\documentclass[final]{revtex4}
\usepackage[dvipdfmx]{graphicx}
\usepackage{amsmath,amssymb}
\usepackage{bm}
\usepackage{here}
\usepackage{hyperref}

\begin{document}

\title{Formation of Twisted Liquid Jets}

\author{Akira Kageyama}
\altaffiliation{Graduate School of System Informatics, Kobe University, Kobe 657-8501, Japan}
\email{kage@port.kobe-u.ac.jp}

\author{Yuna Goto}
\altaffiliation{Hakuryo High School, Takasago 676-0827, Hyogo, Japan}

\begin{abstract}
Liquid jets issued from a non-circular orifice exhibit oscillation owing to the surface tension.
When the orifice has an $n$-fold rotational symmetry, a material cross section of the jet interchanges two symmetric shapes alternately.
This oscillation, called axis switching, is a superposition of two ripples oppositely propagating in the azimuthal direction around the axis.
In this study, we used computer simulations to demonstrate that we can pick up one of the two ripples by adjusting the initial velocity profile of the orifice.
As a result of the single wave propagation in the azimuth, the jet surface shows a twisted appearance.
In contrast to the swirling jets, the twisted jet has no angular momentum around the axis.
We numerically demonstrated the formation of twisted jets with various cross sections, including a regular square.
\end{abstract}

\maketitle

\section{Introduction}
A liquid jet issued from an elliptical orifice exhibits oscillation, which is known as axis switching.
The major and minor axes of the jet's cross section interchange alternately~\cite{Rayleigh1879,Taylor1960,Eggers2008}.
The axis switching of elliptical jets have been studied extensively using theory~\cite{Bechtel1989,Bechtel1998,Amini2011,Pitrou2018c},
experiments~\cite{Kasyap2009,Jaberi2019b,Gu2017},
and a combination of both~\cite{Amini2014,Gu2017}.
For a co-moving observer with the fluid, the axis switching appears as a standing wave of the boundary curve of the cross section~\cite{Tadjfar2019}.
The surface tension with the curvature of the boundary is the driving force of the wave.

The axis switching of elliptical jets is a special case of the standing wave in the non-circular cross section with an $n$-fold rotational symmetry~\cite{Rayleigh1879,Geer1983,Gutmark1999,Rajesh2016,Jaberi2019b}.
The jets with other cross sections with $n>2$ show similar successive alternations between two symmetric states.

We can regard the axis switching as a superposition of two symmetric waves that are oppositely propagating in the azimuth, as in the general standing waves.
In this paper, we show that we can extract one of the two waves, when the initial velocity at the orifice is appropriately adjusted.
A single wave that propagates in an azimuthal direction causes a twisted appearance of the jet's surface, or formation of the twisted prism jet.

It is established that a liquid jet has a helical shape when the fluid is rotating around the central axis~\cite{Ponstein1959,Caulk1979a,Billant1998,Kubitschek2007,Siamas2009,Wang2018}.
In contrast to the swirling jets, the twisted jets studied in this paper have no angular momentum around the axes.
We focus on the early stages of the oscillation for just a couple of cycles after exiting from the orifice, while lots of recent studies on the liquid jet oscillations in the literature focus on the instabilities and breakups~\cite{Pimbley1976,Lin2003,Park2006a,Eggers2008,Liu2008a,Kasyap2009,Amini2012,Amini2014,Wang2015,Morad2020a,Morad2020b}.

\section{Method}

We considered an inviscid fluid issued from a non-circular aperture with an $n$-fold rotational symmetry.
We ignored the spatial derivatives of physical variables along the jet.
Assuming a uniform velocity along the jet, the fluid elements left from the orifice at the same time remain in a common plane.
The boundary shape changes due to the surface tension and the curvature in the plane.
We ignored the gravity and pressure gradient in the cross section, when compared to surface tension.

\subsection{Deep water approximation} \label{190504170831}

We first re-derive the dispersion relation of the surface oscillation for a jet issued from a non-circular orifice with an $n$-fold rotational symmetry.
The area of the orifice is $\pi r_0^2$, where $r_0$ is the radius of the reference circle.
We consider a two-dimensional dynamics in a material cross section $\Lambda$ of the jet and its circumference $\Gamma$.
The radius $r_s$ of $\Gamma$ at a time $t$ in the cylindrical coordinate system is given by 
\begin{equation} \label{190424180329}
  r_s(\vartheta,t) = r_0 \left\{1 + \delta(t) \cos n\vartheta - \delta(t)^2/4\right\}
\end{equation}
where $\delta$ is a small (nondimensional) amplitude compared with the reference circle.
The last correction term $\delta^2/4$ is necessary to keep the $\Lambda$'s area constant ($=\pi r_0^2)$ to $O(\delta^2)$.
We assume that the velocity $\bm{v}$ in $\Lambda$ is a potential flow, 
$\bm{v} = \nabla \psi$,
with 
\begin{equation}  \label{200216115100} 
  \psi = c\,  r^n \cos n\vartheta,
\end{equation} 
where $c$ is a constant.
The kinetic energy $\mathcal{K}$ is
\begin{equation}  \label{200221105343} 
  \mathcal{K} = \frac{\rho}{2} \int_\Lambda \bm{v}^2\, \mathrm{d}S =\frac{\rho}{2} \oint_\Gamma \psi(r_s) v_s \,\mathrm{d} \ell,
\end{equation} 
where $\rho$ is the fluid mass density and $v_s$ is the radial velocity on $\Gamma$.
The line element $\mathrm{d} \ell$ is, from eq.~\eqref{190424180329},
\begin{align}  
  \mathrm{d}\ell &= \sqrt{\mathrm{d}r_s^2 + r_s^2\,\mathrm{d}\vartheta^2} \label{200220133242}\\
                         &= r_0\, \left\{
  							1 + \delta\, \cos n\vartheta
        				 			+ \frac{\delta^2}{2}  
				   						\left(
						  					n^2 \sin^2 n\vartheta - \frac{1}{2}
										\right)
					\right\} \mathrm{d}\vartheta, \label{200220133251}
\end{align} 
to $O(\delta^2)$.
As both $v_s$ and $\psi$ are of $O(\delta)$, 
only the zero-th order term in eq.~\eqref{200220133251} is enough to calculate $\mathcal{K}$ to $O(\delta^2)$.
In other words, we can take the reference circle $r=r_0$ for finding the integral in eq.~\eqref{200221105343}.
The coefficient $c$ in eq.~\eqref{200216115100} is determined by the boundary condition of $\psi$ on $r=r_0$.
Using the time derivative of eq.~\eqref{190424180329} to $O(\delta)$, we get
\begin{equation}  \label{200221131737} 
  v_0 = \partial_r \psi(r_0)  = r_0 \, \dot{\delta}\cos(n\vartheta),
\end{equation} 
or
\begin{equation}  \label{200221141410} 
  c = \frac{\dot{\delta}}{n\, r_0^{n-2}}.
\end{equation} 
Therefore, the total kinetic energy is given by
\begin{equation}  \label{1904270000e}
  \mathcal{K} = \frac{\rho}{2} \oint \psi(r_0)\, v_0 \, r_0\,\mathrm{d}\vartheta   
     = \frac{\rho \pi r_0^4}{2n} \dot{\delta}^2.
\end{equation} 
%
On the other hand, we have to take the $O(\delta^2)$ term in eq.~\eqref{200220133251} into account in the line integral for finding the potential energy by the surface tension $\sigma$;
\begin{align}
  \mathcal{U} &= \sigma \oint_\Gamma \mathrm{d}\ell   \label{190427110000a} \\
     &= 2\pi\sigma r_0 + \sigma r_0 \frac{\pi(n^2-1)}{2} \delta^2. \label{190427110000c}
\end{align}
Combining eqs.~\eqref{1904270000e} and~\eqref{190427110000c}, 
we get the Lagrangian of the system;
\begin{equation}  \label{190427112229} 
  \mathcal{L}(\delta,\dot\delta) = \frac{\rho \pi r_0^4}{2n} \dot{\delta}^2 - \frac{\sigma r_0 \pi (n^2-1)}{2} \delta^2.
\end{equation} 
The solution of the equation of motion shows a harmonic oscillation with the angular frequency
\begin{equation}  \label{190427113246} 
  \omega = \tau_\sigma^{-1}\,\sqrt{n(n^2-1)},
\end{equation} 
with
\begin{equation}  \label{200510100635} 
  \tau_\sigma = \sqrt{\rho r_0^3/\sigma}.
\end{equation} 
The dispersion relation~\eqref{190427113246} was theoretically derived by Rayleigh~\cite{Rayleigh1879} and was experimentally confirmed for $n=2$, $3$, and $4$ by him, as well as in other recent experiments~\cite{Gutmark1999,Wang2015,Rajesh2016}.

For experiments, spatial wavelength $\lambda$ of the oscillation along the jet as a function of Weber number $\mathrm{We}$
is more convenient than the temporal relation~\eqref{190427113246};
\begin{equation}  \label{200511201331} 
  \mathrm{We} \equiv \rho U^2 r_0 / \sigma = (U/u_0)^2,
\end{equation} 
where $U$ and $u_0$ are jet velocity and characteristic velocity $u_0\equiv r_0 / \tau_\sigma$, respectively.
Since $\lambda = 2\pi U / \omega$, eq.~\eqref{190427113246} means
\begin{equation}  \label{200511200601} 
  \frac{\lambda}{r_0} = \frac{2\pi\, \sqrt{\mathrm{We}}}{\sqrt{n(n^2-1)}}.
\end{equation} 
This relation is confirmed by experiments~\cite{Amini2012} and recent 3-dimensional numerical simulations~\cite{Morad2020a,Morad2020b}.

The $r^n$ dependency of $\psi$ in eq.~\eqref{200216115100}, or $r^{n-1}$ dependency of $\bm{v}$ means that the flow is almost absent near the center $r=0$ for $n \ge 3$, as given in the deep-water theory.
We assume that the flow is localized below the surface in a layer with a constant width $w_s$ for $n\ge 3$.
The kinetic energy to $O(\delta^2)$ owing to the localized flow in the layer is approximated by
\begin{equation}  \label{190427170851} 
  \mathcal{K} = \frac{\rho}{2} w_s \oint_\Gamma v_0^2\, \mathrm{d}\ell = \frac{\rho \pi r_0^3 w_s}{2} \dot{\delta}^2.
\end{equation} 
Comparing eqs.~\eqref{1904270000e} and~\eqref{190427170851}, we obtain the width
\begin{equation}  \label{190427171013} 
  w_s = \frac{r_0}{n},
\end{equation} 
is inversely proportional to the azimuthal mode number $n$.
The mass in the layer is estimated as
\begin{equation}  \label{190427174520} 
  M' = 2\pi r_0 \rho w_s  = 2M/n,
\end{equation} 
where $M$ is total mass in the cross section; $M = \rho \pi r_0^2$.
We assume that eq.~\eqref{190427174520} is also valid for $n=2$.
This is consistent with the fact that the flow for $n=2$ spans the full radius ($v\propto r$). Therefore, the mass in the ``layer'' should be $M'=M$.

By ignoring the pressure gradient in the cross section $\Lambda$,
we numerically solve the motion of $\Gamma$ by discretizing it as an $N$-polygon and allocating point masses on the vertices.
The surface tension and the pressure forces act on the $N$ particles which carry the fluid momentum in the layer $w_s$.
Although the potential flow is assumed in the above derivation of eq.~\eqref{190427113246},
we apply this simulation method, Surface Point Method, not only to the irrotational flows, but also to the general flows.

\subsection{Discretization and forces}
Let $\bm{x}_i$ and $\dot{\bm{x}}_i$ be the position and velocity of the $i$-th particle $(1\le i \le N)$ on $\Gamma$.
The momentum of the particle is given by
\begin{equation} \label{190323185914}
  \bm{p}_i = m_\mathrm{e} \, \dot{\bm{x}}_i 
\end{equation}
where $m_\mathrm{e}$ is the ``effective'' mass of the particles.
According to eq.~\eqref{190427174520},  it is given by
\begin{equation}  \label{190427175200} 
  m_\mathrm{e} = 2 m_0 /n \quad (n\ge 2),
\end{equation} 
where $m_0 = M/N$

We define three kinds of tangential vectors $\bm{t}_{i+}$,  $\bm{t}_{i-}$,  and $\bm{t}_{i}$ as
\begin{equation} \label{190425190518}
  \bm{t}_{i+} = \bm{x}_{i+1} - \bm{x}_{i},
\end{equation}
\begin{equation} \label{190425190545}
  \bm{t}_{i-} = \bm{x}_{i} - \bm{x}_{i-1},
\end{equation}
and
\begin{equation} \label{190323190928}
  \bm{t}_i = \frac{\bm{t}_{i+} + \bm{t}_{i-}}{2}=\frac{\bm{x}_{i+1} - \bm{x}_{i-1}}{2}.
\end{equation}
We use the hat symbol to denote the unit vectors 
such as $\hat{\bm{t}}_{i+} = \bm{t}_{i+}/|\bm{t}_{i+}|$.
The unit normal vector in the $\Lambda$ plane is given by
\begin{equation} \label{190323191003}
  \hat{\bm{n}}_i = \hat{\bm{t}}_i\times\hat{\bm{e}}_z,
\end{equation}
where $\hat{\bm{e}}_z$ is unit vector perpendicular to $\Lambda$.

It is simple to calculate the surface tension force $\bm{F}^s_i$ on $\bm{x}_i$ as;
\begin{equation} \label{190323180117}
  \bm{F}^s_i = \sigma \, ( \hat{\bm{t}}_{i+} - \hat{\bm{t}}_{i-} ).
\end{equation}
The pressure force $\bm{F}^p_i$ acting on the line segment $|\bm{t}_i|$ for the particle $\bm{x}_i$ is given by
\begin{equation} \label{190323175150}
  \bm{F}^p_i = \left(p-p_0\right) \hat{\bm{n}}_i,
\end{equation}
where $p$ and $p_0$ are the internal pressure in the $\Lambda$ and the external atmospheric pressure.
They satisfy the Laplace relation,
\begin{equation} \label{190324121920}
  p(\ell) - p_0 = k(\ell)\sigma,
\end{equation}
where $k$ is the curvature, and $\ell$ is the length along $\Gamma$, measured from any reference point.
We assume a constant pressure given by the following average
\begin{equation} \label{190324122458}
  p = p_0 + \frac{1}{L}\oint k(\ell)\sigma\, \mathrm{d}\ell =  p_0 +\frac{2\pi\sigma}{L},
\end{equation}
where $L$ is the total length of $\Gamma$.
To maintain the incompressibility (area conservation) of the jet,
we assume the following polytropic relation
\begin{equation} \label{190324122926}
  p(t) = p_0 + \frac{2\pi\sigma}{L(t)}\left\{\frac{A_0}{A(t)}\right\}^\gamma,
\end{equation}
where $L(t)$ is the $\Gamma$'s length at time $t$, and $A(t)$ is the $\Lambda$'s area whose initial value (orifice's area) is $A_0$.
The $\gamma$ is an arbitrary large number; we set $\gamma=1000$ for this study.

\subsection{Basic equations of the Surface Point Method}

\begin{figure}[H]   \centering   
  \includegraphics[%
     height=0.5\textheight,%
       width=0.5\hsize,keepaspectratio]%
         {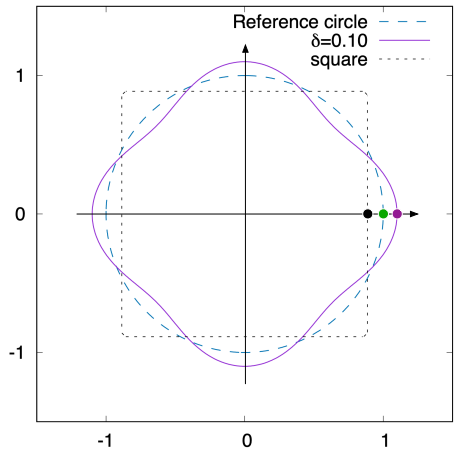}   
      \caption{
      Sample initial cross sections with the same area $A_0=\pi r_0^2$.
      In the simulations, we set $r_0=0.01$~(m).
      The length is normalized by $r_0$.
      The dashed green curve is the reference circle with radius $r_0$.
      The dotted line denotes a regular square that is used in the simulations in Section~\ref{200226145714}.
      The purple curve is an example ($n=4$) of trigonometric profiles with $\delta=0.1$ simulated in Section~\ref{200226145319}.}
      \label{190427145720}
\end{figure}

To summarize, we solve the following equations for a cross section $\Lambda$ with an $n$-fold rotational symmetry,
\begin{align}
  \frac{\mathrm{d} \bm{p}_i}{\mathrm{d}t} &= \bm{F}^p_i + \bm{F}^s_i,  \label{1903231917a}  \\[0.5em]
  \frac{\mathrm{d} \bm{x}_i}{\mathrm{d}t} &= \bm{p}_i  / m_\mathrm{e},  \label{1903231917b} 
\end{align}
where $\bm{F}^s_i$ is the surface tension;
\begin{equation} \label{190323192909}
  \bm{F}^s_i =  \sigma\, ( \hat{\bm{t}}_{i+} - \hat{\bm{t}}_{i-} ),
\end{equation}
and $\bm{F}^p_i $ is the pressure difference force
\begin{equation} \label{190323192857}
  \bm{F}^p_i = \frac{2\pi\sigma}{L(t)}\left\{\frac{A_0}{A(t)}\right\}^\gamma\, \hat{\bm{n}}_i .
\end{equation}
$L(t)$ and $A(t)$ are the length of $\Gamma$ and the area of $\Lambda$ that are respectively calculated by
\begin{equation} \label{190425184805}
  L(t) = \sum_{i=1}^N |\bm{t}_i|,
\end{equation}
and
\begin{equation} \label{190425185144}
  A(t) = \frac{1}{2}\sum_{i=1}^N \bm{x}_i\cdot\hat{\bm{n}}_i\, |\bm{t}_i|.
\end{equation}
In the following simulations, we consider jets with $A_0=\pi r_0^2$ with $r_0=0.01$~(m).
The particle number $N$ is $100$ in all the simulations.
The mass density and the surface tension are $\rho= 998$ (kg/m$^3$) and $\sigma=73\times 10^{-3}$ (N/m), respectively.
Time and length will be presented in non-dimensional values normalized by $\tau_\sigma=1.17\times 10^{-1} \, \mathrm{(s)}$ and $r_0$.
The dashed green circle in Fig.~\ref{190427145720} depicts the reference circle of radius $r_0$.
The purple curve is an example of the trigonometric profile of eq.~\eqref{190424180329} with $n=4$ and $\delta=0.1$.

During the simulation, the spacing between the neighboring particles on $\Gamma$ may become non-uniform due to the imbalance of the tangential inertia of the particles.
To fix the non-uniformity, we apply the following re-distribution procedure:
We select three consecutive particles, $P_{i-1}$, $P_i$, and $P_{i+1}$, in the $N$ particles.
Then, we calculate a circle that passes through the three particles.
We shift the middle particle $P_i$ along the arc $P_{i-1}P_{i+1}$ so that it is located just on the middle point of the arc.
Note that the amplitude of the surface tension force $\bm{F}_s$ acting on $P_i$, which is inversely proportional to the radius of the local arc, does not change in this shift.

We apply the above procedure consecutively for every three triplet $\{i+1, i, i-1\}$ for all $1\le i \le N$ in one turn of the redistribution procedure.
(In the simulation, the particle $P_0$ is identical to $P_N$, and so $P_{N+1}=P_1$.)
We repeat the turns until the spacing $\Delta_i$ between the particle pairs $P_{i}$ and $P_{i-1}$ is sufficiently uniform.
The criterion for the uniformity is that $\mathrm{max}(\Delta_i) - \mathrm{min}(\Delta_i) < 0.01\times \triangle L$,
where $\triangle L$ is the average spacing; $\triangle L=L/N$.
We have found just one turn for each time step in the following simulations.

\begin{figure}[H]   \centering
  \includegraphics[%
     height=0.6\textheight,%
       width=0.45\hsize,keepaspectratio]%
         {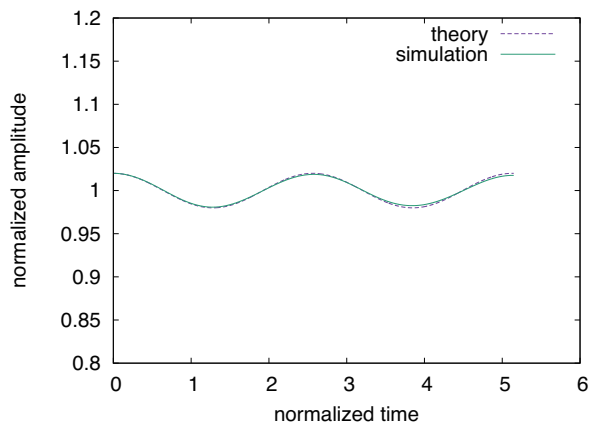}
      \caption{Time development of the surface on the polar axis ($\vartheta=0$) for a small amplitude oscillation $\delta=0.02$ with $n=2$.
      The horizontal axis represents time normalized by $\tau_\sigma$ and the vertical axis represents the oscillation amplitude normalized by $r_0$.
      }
      \label{190505154818}
\end{figure}

\section{Validation}\label{191102162324}

To validate the Surface Point Method, we compare the simulations with the Rayleigh's theory.
A trigonometric orifice with the azimuthal mode number $n$ is taken.
The initial profile of the particles is given by eq.~\eqref{190424180329}.
The initial velocity of each particle is zero.

\begin{figure}[H]   \centering   
  \includegraphics[%
     height=0.6\textheight,%
       width=0.45\hsize,keepaspectratio]%
         {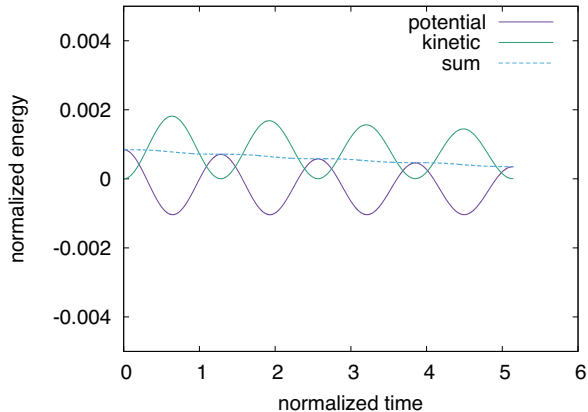}   
      \caption{
      Time development of potential energy, kinetic energy, and their sum
      in the simulation for $\delta=0.02$ with $n=2$.
      Time is normalized by $\tau_\sigma$ and the energy is normalized by $ r_0\sigma$.
      }
      \label{200511162216}
\end{figure} 
Fig.~\ref{190505154818} shows the oscillation for the mode $n=2$ with the initial (nondimensional) amplitude $\delta=0.02$.
The horizontal axis is time (second) and the vertical axis (meter) is position of the surface that crosses the polar axis ($\vartheta=0$).
The two curves obtained by theory (dashed purple) and simulation (solid green) are in good agreement.
This figure shows the time span for two cycles.
The oscillation frequency estimated by the simulation for the first cycle is 
$\omega_{\,\mathrm{s}}=2.44 \, \tau_\sigma^{-1}$,
which coincides with the theoretical value 
$\omega_{\,\mathrm{t}} = 2.44\, \tau_\sigma^{-1}$
given by eq.~\eqref{190427113246}.

To visualize the energy conversion between the potential energy $\mathcal{U} \equiv \sigma\, (\sum_i |\bm{t}_i|-2\pi\,r_0)$ and 
the kinetic energy $\mathcal{K} \equiv \sum_i \bm{p}_i^2/(2m_e)$,
we plot in Fig.~\ref{200511162216} time developments of the energies and their sum.
During the first cycle of the oscillation ($0\le t \le \tau \sim 2.57$ in normalized time), 
the energy conversion is observed for two times.
This reflects two symmetric states in the axis-switching of the jet.
The first conversion from $\mathcal{U}$ to $\mathcal{K}$ ($0\le t \le \tau/4$) is driven by the cross section of horizontally long oval.
The second conversion from $\mathcal{U}$ to $\mathcal{K}$ ($\tau/2 \le t \le 3\tau/4$) is driven by vertically long oval.
In the recent 3-dimensional direct numerical simulation of the elliptic~\cite{Morad2020a} and rectangular~\cite{Morad2020b} cross sections,
the symmetric energy conversions as well as the pressure distribution in the cross section are analyzed in detail.
The gradual decay of the total energy (dashed line in Fig.~\ref{200511162216}) is due to the numerical error of Surface Point Method.

\begin{figure}[H]   \centering   
  \includegraphics[%
     height=0.6\textheight,%
       width=0.45\hsize,keepaspectratio]%
         {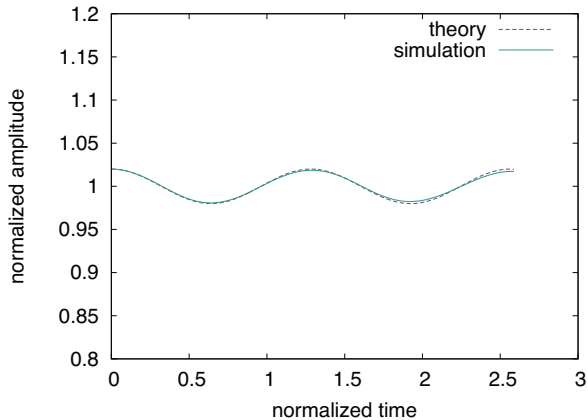}   
      \caption{Same as Fig.~\ref{190505154818}, except that $n=3$.}
      \label{190505172436}
\end{figure} 
Fig.~\ref{190505172436} shows the oscillation for the mode $n=3$.
Other parameters such as $\delta$ are the same as those in Fig.~\ref{190505154818}.
Again, the simulation result is in good agreement with the theory.
The estimated frequency from the simulation is 
$\omega_{\,\mathrm{s}} = 4.88\, \tau_\sigma^{-1}$,
while the theoretical value is 
$\omega_{\,\mathrm{t}} = 4.90\, \tau_\sigma^{-1}$.

\begin{figure}[H]   \centering   
  \includegraphics[%
     height=0.6\textheight,%
       width=0.45\hsize,keepaspectratio]%
         {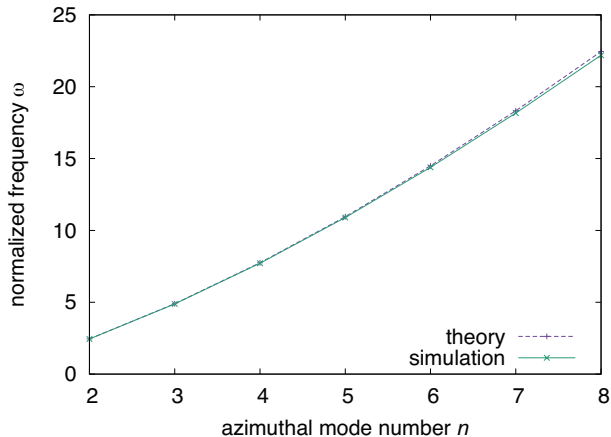}   
      \caption{
      Non-dimensional angular frequency, $\omega$ normalized by $\tau_\sigma^{-1}$,
      as a function of the azimuthal mode number $n$. 
      Simulation data (solid green) are taken from the trigonometric profiles with $\delta=0.02$.
      The theoretical data (dashed purple) are given by eq.~\eqref{190427113246}.}
      \label{190505230548}
\end{figure}

To confirm the mode number dependency, $n(n^2-1)$, in the frequency $\omega$ [eq.~\eqref{190427113246}],
we performed other linear simulations with different $n$ values.
The results are summarized in Fig.~\ref{190505230548}.
The Surface Point Method successfully reproduces the dispersion relation.

\begin{figure}[H]   \centering   
  \includegraphics[%
     height=0.5\textheight,%
       width=0.5\hsize,keepaspectratio]%
         {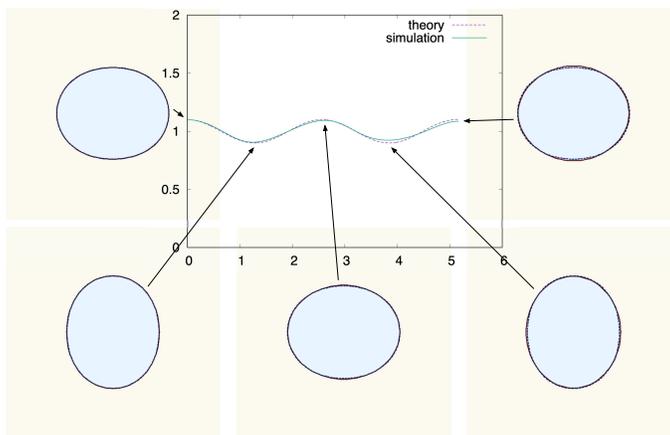}   
      \caption{Graph in the upper middle section is the time development of oscillation amplitude 
      normalized by $r_0$
      for the trigonometric jet with $\delta=0.1$ and $n=2$,
      for theory (dashed purple) and simulation (solid green).
      The snapshots shown around the graph are the cross sections of the jet at the designated times, for theory (dashed line) and simulation (solid line).
      The theoretical and simulated profiles almost overlap completely.
      }
      \label{190506185928}
\end{figure} 

Now, we apply the simulation for the larger amplitudes; $\delta=0.1$.
The graph in the upper middle in Fig~\ref{190506185928} is the same as Fig.~\ref{190505154818};
the time development of the oscillation amplitude for $n=2$ is given by the dashed curve (theory) and solid curve (simulation).
The oscillation frequency estimated from the simulation is 
$\omega_{\,\mathrm{s}}=2.40\, \tau_\sigma^{-1}$.
Compared with the theoretical value
 [$\omega_{\,\mathrm{t}}=2.44\, \tau_\sigma^{-1}$], 
 the relative error is $1.9$~\%.
We placed five snapshots of the simulated cross section $\Lambda$ at designated times by arrows around the graph.
In these cross sections, the jet surfaces obtained by the simulation and theory are denoted by the solid and dashed curves, respectively.
They overlap almost completely in this case.

\begin{figure}[H]   \centering   
  \includegraphics[%
     height=0.6\textheight,%
       width=0.5\hsize,keepaspectratio]%
         {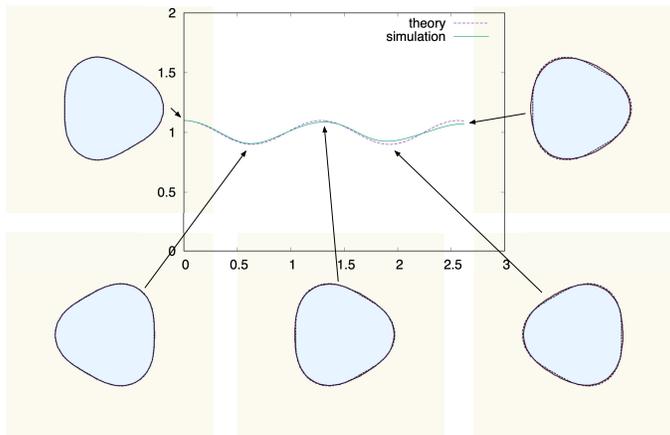}   
      \caption{Same as Fig.~\ref{190506185928}, but with the mode number $n=3$.}
      \label{190506204937}
\end{figure} 
Fig.~\ref{190506204937} is the same as Fig.~\ref{190506185928} except the mode number $n=3$.
The oscillation frequency estimated from the simulation is 
$\omega_{\,\mathrm{s}}=4.79\, \tau_\sigma^{-1}$.
The relative error compared with the theory 
[$\omega_{\,\mathrm{t}}=4.90\, \tau_\sigma^{-1}$]
 is $2.1$~\%.

The above simulations suggest that the Surface Point Method can predict the frequency with an accuracy of about $2$~\% as long as the oscillation amplitude $\delta \le 0.1$, at least during the first cycle.

\section{Twisted square jet}\label{200226145714}
Here, we consider an orifice of regular square with area $\pi r_0^2$, which is shown by dotted lines in Fig.~\ref{190427145720}.
The ``amplitude'' of the regular square is comparable to the trigonometric profile with $n=4$ and $\delta=0.1$ on the polar axis, as indicated by black, green, and purple dots in Fig.~\ref{190427145720}.

In simulations described in the previous section,
the surface particles are all stationary (no velocity) at $t=0$ before moving by the imbalance between the surface tension and the pressure forces.
As we solve the time development of each particle by eqs.~\eqref{1903231917a} and~\eqref{1903231917b} in Surface Point Method.
It is possible to specify a non-zero velocity profile in the initial condition.
We show in this section that shape of the surface $\Gamma$ exhibits rotation for some initial velocities.

\begin{figure}[H]   \centering   
  \includegraphics[%
     height=0.4\textheight,%
       width=0.4\hsize,keepaspectratio]%
         {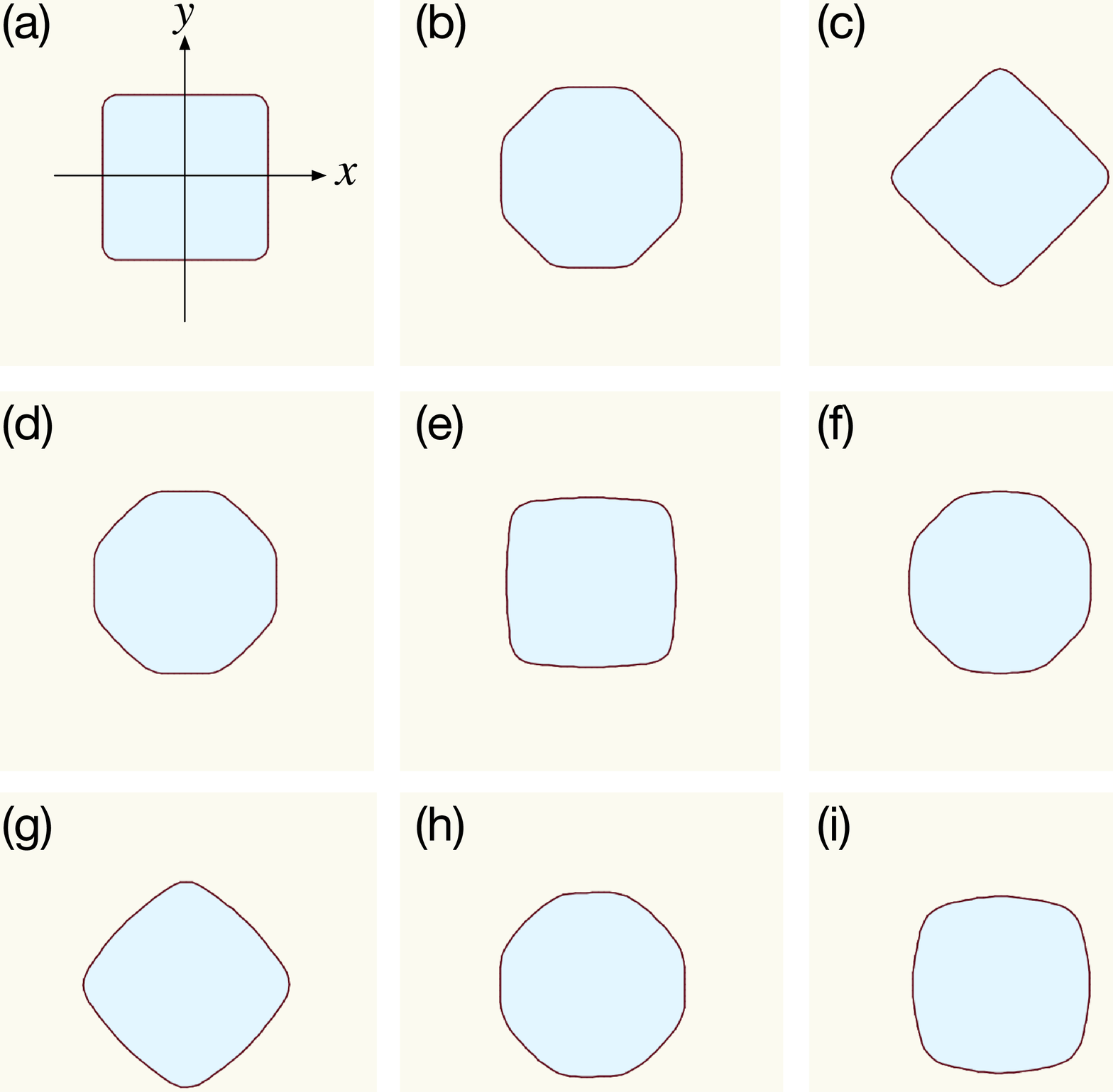}   
      \caption{Cross sections of the regular square jet.
      The panels (a) to~(i) are a time sequence of snapshots with constant intervals.
      The velocity of each particle in the initial condition is absent.}
      \label{191105093532}
\end{figure} 
Before showing the rotation, we first present a standard oscillation or axis switching of the regular square in Fig.~\ref{191105093532}.
The four edges in the initial condition, shown in Fig.~\ref{191105093532}(a), are parallel to the lines $x=\pm 1$ and $y=\pm 1$.
(Here, $x$ and $y$ axes are temporarily defined for convenience of explanation.)
We slightly rounded the four corners of the square to avoid singular tension on the vertices.
The velocity of each particle on the surface is zero in the initial condition.
The panels~(a) to~(i) in Fig.~\ref{191105093532} are the snapshots taken with a constant interval.
The interval is one-fourth of the period 
$\tau= 0.844\, \tau_\sigma$, 
which is measured by the surface position on the polar axis ($+x$-axis).
To elaborate, we monitor the $x$ coordinate, $x_\text{cross}$, of the surface $\Gamma$ that crosses the polar axis and measure the time for one period of the oscillation;
the initial minimum of $x_\text{cross}$ is in Fig.~\ref{191105093532}(a), and $x_\text{cross}$ increases to its maximum value in Fig.~\ref{191105093532}(c);
then it gets back to the minimum in Fig.~\ref{191105093532}(e).
The period $\tau$ is defined by the time from Fig.~\ref{191105093532}(a) to Fig.~\ref{191105093532}(e).
Incidentally, from the Rayleigh's theory [eq.~\eqref{190427113246}], the period is found to be 
$0.811\, \tau_\sigma$
for $n=4$.

As shown in the previous studies~\cite{Geer1983,Wang2015,Rajesh2016},
the profile of the cross section demonstrates a periodic deformation and interchanging of the two configurations with the same shape but different angles.
The surface tension pulls the particles near the four vertices in the initial configuration [Fig.~\ref{191105093532}(a)].
The pulled particles form four new edges that are parallel to $x+y=\pm 1$ and $x-y=\pm 1$.
The profile of the cross section at $t=\tau/4$ is close to a regular octagon [Fig.~\ref{191105093532}(b)].
The length of the newly born edges grow until it thoroughly erodes the original edges ($x=\pm 1$ and $y=\pm 1$).
The profile then becomes a regular square with edges parallel to $x+y=\pm 1$ and $x-y=\pm 1$ [Fig.~\ref{191105093532}(c)].
The transition from the initial square [Fig.~\ref{191105093532}(a)] to the rotated square [Fig.~\ref{191105093532}(c)] extends similarly from Fig.~\ref{191105093532}(d) to Fig.~\ref{191105093532}(g), and thereafter.
This alternate transition of the two configurations is the axis switching in the $4$-fold rotational symmetry.

\begin{figure}[H]   \centering   
  \includegraphics[%
     height=0.5\textheight,%
       width=0.5\hsize,keepaspectratio]%
         {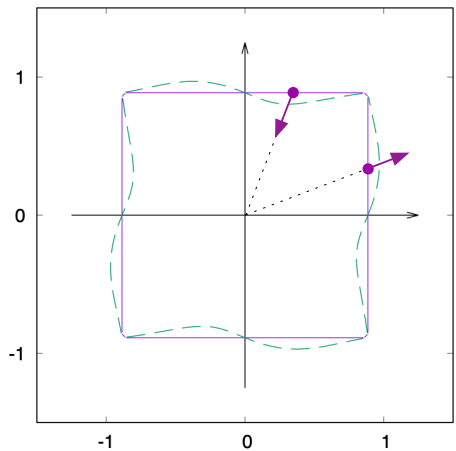}   
      \caption{Profile of the square jet (solid purple) and the initial velocity profile on the surface (dashed green) used in the simulation.
      The initial profile (normalized by $r_0$) is a regular square with the same area as that of the reference circle $\pi r_0^2$.
      The velocity has only the radial component as depicted by the arrows.}
      \label{191120173337}
\end{figure} 
We now set non-zero velocity to the surface particles on the regular square in the initial condition. 
Velocity of the $i$-th particle in the cylindrical coordinate system is given by,
\begin{equation}  \label{191107105446} 
  \bm{v}^i = (v_r^i, v_\vartheta^i) = (u \sin(4\vartheta_i), 0),
\end{equation} 
with $u=0.894\, u_0$.
The profile is shown in Fig.~\ref{191120173337}.
The velocity $u$ pushes the vertical edge ($x=\text{const.}$) in the first quadrant to the outward direction,
while it pulls the horizontal edge ($y=\text{const.}$) to the inward direction in the same quadrant.

\begin{figure}[H]   \centering   
  \includegraphics[%
     height=0.4\textheight,%
       width=0.4\hsize,keepaspectratio]%
         {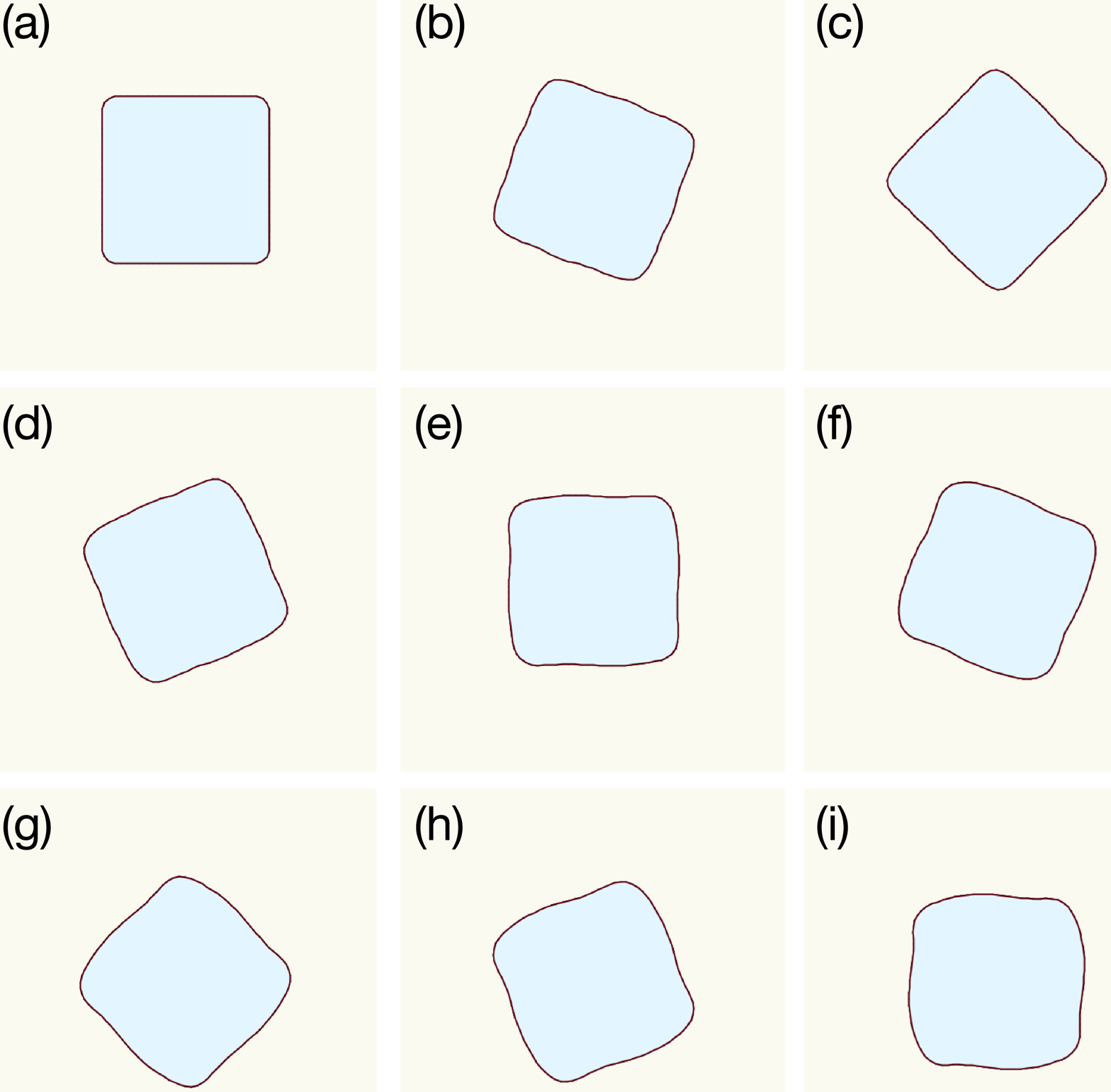}   
      \caption{Cross sections of the regular square jet.
      The panels (a) to~(i) are a time sequence of the snapshots with constant intervals.
      Each particle on the surface of the cross section moves mainly in the radial direction.
      The rotating appearance comes from the ripple propagating on the surface in a clockwise direction.}
      \label{191113194202}
\end{figure}

Fig.~\ref{191113194202} shows a sequence of cross sections taken by the same procedure as in Fig.~\ref{191105093532}:
We first defined the period $\tau$ of the oscillation by monitoring the coordinate $x_\text{cross}$ of the surface that crosses the $+x$-axis,
from the initial minimum [Fig.~\ref{191113194202}(a)] to the next minimum [Fig.~\ref{191113194202}(e)] 
and have found that 
$\tau = 0.860\, \tau_\sigma$.
We took snapshots of the cross sections with constant interval of a forth of $\tau$ from Fig.~\ref{191113194202}(a) to Fig.~\ref{191113194202}(i), for $2\tau$.

During the time span, the square profile rotates for $180^\circ$.
However, the apparent rotation of the surface curve does not mean the actual rotation of the surface particles around the center,
in contrast to the swirling jets~\cite{Ponstein1959,Caulk1979a,Billant1998,Kubitschek2007,Siamas2009,Wang2018}.
In fact, each of the surface particle moves almost only in the radial direction.
The absence of swirling is natural, as the total angular momentum of the surface particles is zero in the initial condition;
the azimuthal velocity $v_\vartheta^i=0$ [see eq.~\eqref{191107105446}].
(Although, the conservation of the angular momentum is slightly violated in the redistribution procedure of the particles.)

The rotation of the square profile is not sensitive to the value of the perpendicular velocity $u$:
The turning also appears when 
$u=0.179\,u_0$, 
which is $20\%$ of the value adopted in the simulation that is shown in Fig.~\ref{191113194202},
although the square profiles are not as sharp as in Fig.~\ref{191113194202}.
The rotation disappears when $u$ is $10\%$, i.e., 
$u=0.0894\, u_0$.

As the sequence from (a) to~(i) in Fig.~\ref{191113194202} indicates,
the initial velocity given by eq.~\eqref{191107105446} leads to the clockwise rotation of the square.
Observing the velocity profile in Fig.~\ref{191120173337}, we can infer the reversed velocity, as follows
\begin{equation}  \label{191107161208} 
    \bm{v}^i = (v_r^i, v_\vartheta^i) = (-u \sin(4\vartheta_i), 0),
\end{equation} 
would lead to the counterclockwise rotation.
We have confirmed that it is the case.

\begin{figure}[H]   \centering   
  \includegraphics[%
     height=0.4\textheight,%
       width=0.4\hsize,keepaspectratio]%
         {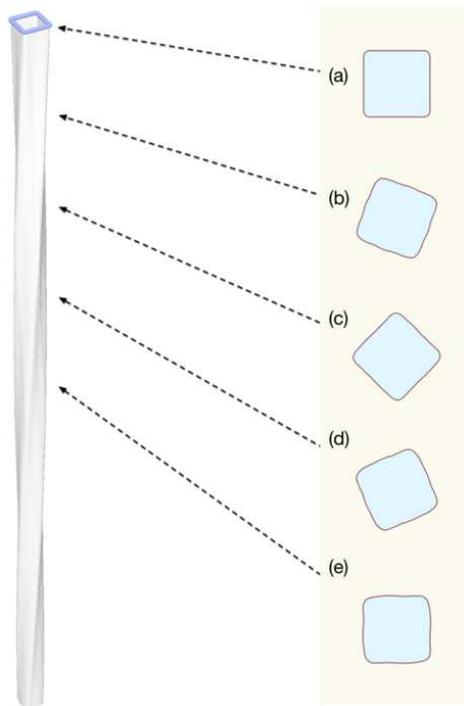}   
      \caption{Twisted square jet obtained by the simulation.
      The normalized flow speed along the jet is assumed be 
      $U=44.7\,u_0$, or $\mathrm{We}=2000$.
      The length of the jet is $77\, r_0$.
      The amplitude of the initial perpendicular velocity 
      $u=0.894\, u_0$.
      The panels (a) to (e) correspond to the cross sections shown in Fig.~\ref{191113194202} with the same labels.}
      \label{191107175004}
\end{figure} 
As the z-component of the velocity $U$ of the jet is supposed to be uniform in the Surface Point Method,
we can construct the three-dimensional surface of the jet from the two-dimensional curves $\Gamma$ in each time.
We put together surface polygons by consecutively connecting the surface points at time $t$ on $z=U \times t$ and points at $t+\triangle t$ on $z=U \times (t+\triangle t)$.
We used the ray tracing software POV-Ray to render the surface of the twisted jet, which was constructed in a perspective view;
see the left part of Fig.~\ref{191107175004}.
In this rendering, we assume 
$U=44.7\,u_0$, or $\mathrm{We}=(U/u_0)^2 = 2000$.
The length of the jet 
$L_z= 77\, r_0$.
The cross section rotates for about $180^\circ$ during this distance.
In the right part of this figure,
panels labeled~(a) to~(e) are the cross sections at the designated location (or time),
which are the same as the ones in Fig.~\ref{191113194202}.

\section{Twisted jets of other cross sections}\label{200226145319}
\begin{figure}[H]   \centering   
  \includegraphics[%
     height=0.6\textheight,%
       width=0.5\hsize,keepaspectratio]%
         {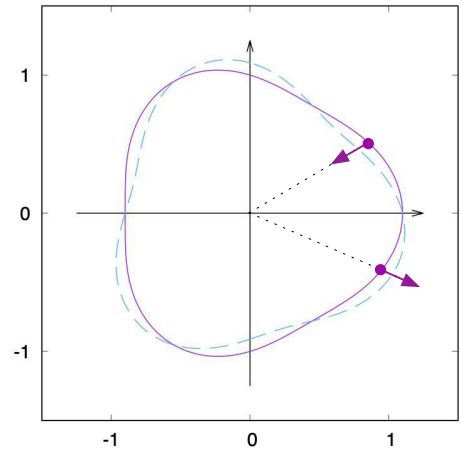}   
      \caption{Profiles of the initial surface (dashed green curve) and the radial velocity (solid purple curve) for 
      the trigonometric profile (normalized by $r_0$)
      with the azimuthal mode number $n=3$, with nondimensional amplitude $\delta=0.1$.}
      \label{191122170447}
\end{figure}

\begin{figure}[H]   \centering   
  \includegraphics[%
     height=0.4\textheight,%
       width=0.4\hsize,keepaspectratio]%
         {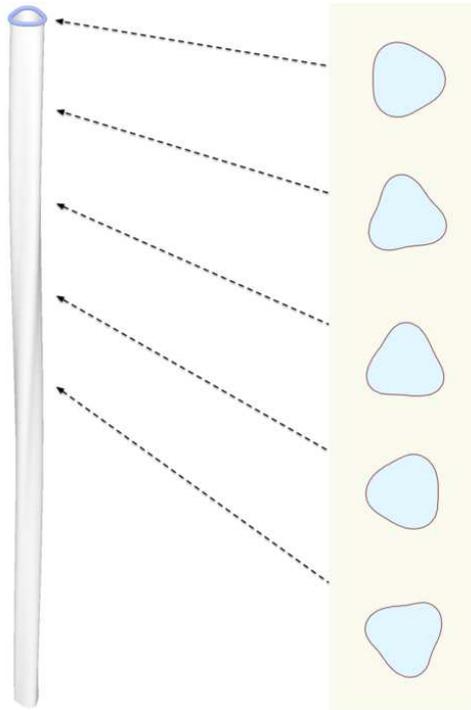}   
      \caption{Twisted prism jet with quasi-triangular cross section.
      The initial profile (the orifice denoted by blue) is given by the trigonometric profile with $n=3$ and $\delta=0.1$.}
      \label{191108162309}
\end{figure}

The twisted square jet described in the previous section was formed by adjusting the azimuthal phase of the $v_r(\vartheta)$ profile in the initial condition.
Similarly, we can construct the twisted jets issued from the $n$-trigonometric orifices given by eq.~\eqref{190424180329}.
For the formation of twisted jets, we set the initial velocity profile as
\begin{equation}  \label{191110180747} 
   \bm{v}^i = (v^i_r, v^i_\vartheta) = (-u \sin(n\vartheta_i), 0),
\end{equation} 
where $u$ is the same value as given in Fig.~\ref{191113194202}; 
$u=0.894\,u_0$.
Fig.~\ref{191122170447} shows the case with $n=3$ and $\delta=0.1$.
The solid purple curve depicts the quasi-triangular profile.
The dashed green curve denotes the profile of $v_r$.
The two arrows in the figure exemplify the velocity on the surface;
the velocities have only the radial component.
Fig.~\ref{191108162309} shows the simulation results.
As in the case of Fig.~\ref{191107175004}, we set 
$U=44.7\, u_0$
for the rendering of the jet with the same length 
$L_z= 77\, r_0$.
The twisted quasi-triangular jet is formed.


\begin{figure}[H]   \centering   
  \includegraphics[%
     height=0.45\textheight,%
       width=0.45\hsize,keepaspectratio]%
         {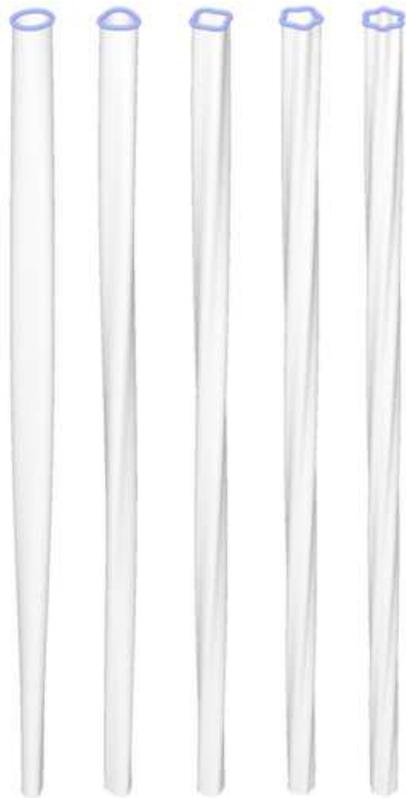}   
      \caption{Twisted jets with $n$-trigonometric cross sections.
      $n=2, 3, 4, 5$, and $6$, from left to right.
      The jet length $L_z=0.77$~(m) and the nondimensional amplitude $\delta=0.1$ are all the same.}
      \label{191114181445}
\end{figure} 
Fig.~\ref{191114181445} compiles the twisted prism jets by $n$-trigonometric orifices with $n=2$, $3$ (same as Fig.~\ref{191108162309}), $4$, $5$, and $6$, from left to right.
The initial perpendicular velocity $u$, parallel velocity $U$, and the jet's length $L_z$ are the same as those given in Fig.~\ref{191108162309}.

\section{Summary}

A liquid jet issued from a non-circular aperture with an $n$-fold rotational symmetry exhibits characteristic surface oscillation called axis-switching.
For an observer moving with the fluid,
the axis-switching is a standing wave of the liquid boundary in a material cross section.

The standing wave is a superposition of two, oppositely propagating, symmetric ripples.
We have shown that we can launch one of the two ripples that propagates in a single azimuthal direction by adjusting the perpendicular velocity profile at the orifice.
The single ripple propagating in an azimuthal direction means the formation of a twisted surface of the jet.
The twisted jet is robust in the sense that it is not very sensitive to the value of the perpendicular velocity amplitude.

The perpendicular velocity on the orifice has only $r$ component in the cylindrical coordinates.
This means that the angular momentum about the jet axis is zero, in contrast to the swirling jets.
This also means that the flow is not a potential flow as the vorticity $\omega_z = - (1/r) (\partial_\vartheta v_r)$ is not zero on the surface.
To realize the twisted jet in the laboratory experiments,
we have to inject $\omega_z$ to the flow before exiting from the orifice.

We have developed a simple two-dimensional simulation model, the Surface Point Method, for the surface oscillation.
Despite its simplicity, this method can successfully simulate the surface oscillation at least for a couple of first cycles.
A limitation of this method is that it cannot simulate complex cross sections with multiple azimuthal modes, as the ``effective mass'' [eq.~\eqref{190427175200}] of each particle implicitly assumes a single mode.
However, this would not be a problem as long as the axis-switching or twisted jet is of interest, as these are observed in cross sections with an $n$-fold rotational symmetry.

The experimental verification of the twisted prism jet is an intriguing challenge. 
In this paper, we have ignored the gravity and the surrounding air. 
When the jet velocity or Weber number is large, the interaction with the surrounding air would not be negligible. 
On the other hand, when the jet velocity is small, the gravity takes effect. 
When a slow jet is ejected in a vertically downward direction, 
the gravity acceleration changes the cross-sectional area of the jet. 
Even in that case, the twisted jet would be observed because gravity plays no role in the horizontal dynamics of a cross-section observed in a frame of reference falling with the cross-section, 
except for the temporal decrease of the cross-sectional area. 
One technical difficulty for experiments would be to impose an appropriate perpendicular velocity profile at the orifice. 
One immediate suggestion is to place small perpendicular vents just before the orifice.

\begin{acknowledgments}
This work was supported by JSPS KAKENHI Grant Number 17H02998 and GSC-ROOT Program.
\end{acknowledgments}

%

\end{document}